# Nanoscale soil-water retention mechanism of unsaturated clay via MD and machine learning

Zhe Zhang, Xiaoyu Song*

*Department of Civil and Coastal Engineering, University of Florida, Gainesville, Florida*

**Abstract**

In this article, we investigate the nanoscale soil-water retention mechanism of unsaturated clay through molecular dynamics and machine learning. Pyrophyllite was chosen due to its stable structure and as the precursor of other 2:1 clay minerals. A series of molecular dynamics simulations of clay at low degrees of saturation were conducted. Soil water was represented by a point cloud through the center-of-mass method. Water-air interface area was measured numerically by the alpha-shape method. The soil-water retention mechanism at the nanoscale was analyzed by distinguishing adsorptive pressure and capillary pressure at different mass water contents and considering the apparent capillary interface area (i.e., water-air interface area per unit water volume). The water number density profile was used to quantify the adsorption effect. A neural-network based machine learning technique was utilized to construct function relationships among matric suction, the mass water content, and the apparent water-air interface area. Our numerical results have demonstrated from a nanoscale perspective that the adsorption effect is dominated by the van der Waals force and hydroxyl hydration between the clay surface and water. As the mass water content increases, the adsorption pressure decreases, and capillarity plays a prominent role in the soil-water retention mechanism at the nanoscale.

# 1. Introduction

The physics and mechanics of unsaturated soils are important in geotechnical and geoenvironmental engineering (e.g., Terzaghi et al., 1996; Gens, 2010; Fredlund, 2006; Ng and Menzies, 2014; Song, 2017; Alonso, 2021; Menon and Song, 2022, 2023). Soil-water retention/characteristic curve (SWRC) is a mathematical relationship between soil water content and matric suction (e.g., Brooks, 1965; Van Genuchten, 1980; Fredlund and Rahardjo, 1993; Fredlund and Xing, 1994; Niu et al., 2020; Cao et al., 2018; Chen et al., 2019). It is a fundamental constitutive law for modeling the physics and mechanics of unsaturated soils. For instance, a soil water retention curve is required in modeling multiphase fluid flow, shear strength, deformation, and stress-strain relationships of unsaturated soils (e.g., Alonso et al., 1990; Wheeler et al., 2003; Macari et al., 2003; Hoyos and Arduino,

*Corresponding author
  Email address: xysong@ufl.edu (Xiaoyu Song)

2008; Alonso et al., 2010). In unsaturated soil mechanics and continuum-based numerical methods for modeling unsaturated soils with no osmosis effect, matric suction is usually assumed to be the difference between pore air pressure and pore water pressure and the latter is usually assumed to be the capillary pressure due to water menisci (e.g., Fredlund and Rahardjo, 1993; Borja, 2004, 2006; Menon and Song, 2020, 2021; Song et al., 2017; Wang and Song, 2020) without considering adsorptive water pressure. The adsorptive water pressure might be ignored at a high degree of saturation. However, at a low degree of saturation, it should be considered to interpret high matric suction (e.g., in the order of hundred megapascals) (Fredlund and Rahardjo, 1993; Lu and Likos, 2006; Zhang and Lu, 2019). It is noted that the pressure of capillary water is lower than air pressure due to the curve water-air interface (i.e., meniscus), and pressure of adsorptive water is higher than air pressure due to adsorptive force (Luo et al., 2022). Furthermore, both experimental and theoretical studies have suggested that the water-air interface should be taken into account to better describe soil water retention curves of unsaturated soils (Fredlund and Morgenstern, 1977; Hassanizadeh and Gray, 1990; Joekar-Niasar et al., 2008; Lourenço et al., 2012; Lu and Likos, 2006; Likos, 2014; Fredlund, 2006). We refer to the related literature for a thermodynamic justification (e.g., Houlsby, 1997; Nikooee et al., 2013) of including the water-air interface in the soil-water retention curve of unsaturated soils. In Fredlund and Morgenstern (1977), the water-air interface was first incorporated into stress analysis of unsaturated soils where the air-water interface is treated as an independent phase. In Lu and Likos (2006), the interfacial effects are lumped into the suction stress in addition to capillary pressure. In Nikooee et al. (2013), the interfacial energy and air-water specific interfacial area are integrated into an effective stress tensor using a thermodynamic approach. Interfacial force arises due to the unbalanced force exerted on two sides of interfaces, which may influence the macroscopic soil behavior. Several physicochemical effects contribute to the origin of interface force, such as van der Waals forces, surface tension, and electric double-layer forces. These interface forces could produce surface energy change and deformation of soil (Butt et al., 2013).

Over the past decades, computational modeling through physics-based numerical methods has gained success in resolving and quantifying water-air interfaces in porous media. One standard method is the pore-network modeling technology (Lowry and Miller, 1995; Joekar-Niasar et al., 2008). Several techniques have also been developed to measure the water-air interface area in porous media (Costanza-Robinson and Brusseau, 2002; Chen and Kibbey, 2006; Wildenschild et al., 2002; Brusseau et al., 2007; Lourenço et al., 2012). However, the configuration of pore network is user-defined instead of the actual pore space in nature. Moreover, it remains challenging to quantify the impact of adsorption on SWRC and explain the mechanism of soil-water adsorption at the nanoscale. At the nanoscale adsorptive forces in fine-grained clay become pronounced and could modify the water structure, e.g., adsorptive water film tightly attached to clay surface (Evans et al., 1986; Tuller et al., 1999). It is noted that the laboratory measurement techniques only suffice to account for capillary effects in the water retention mechanism (Likos et al., 2019). Indeed no viable experimental technique exists to quantify adsorption and its impact on SWRC in unsaturated soils at the nanoscale (Lu, 2016).

As a numerical method at the atomic scale, molecular dynamics (MD) can naturally consider adsorption at the nanoscale. With advances in high-performance supercomputers, MD simulations have been extensively used to gain detailed insights into the physics and mechanics of unsaturated soils at the atomistic scale (e.g., Cygan et al., 2004; Katti et al., 2015; Song and Zhang, 2021; Song et al., 2018; Song and Wang, 2019). MD is a computational simulation technique that numerically solves Newton's equations of a classical $N$-body system at equilibrium (Frenkel and Smit, 2001; Allen and Tildesley, 2017; Plimpton, 1995). It is a viable numerical tool to study the effect of soil-water interactions on the physics and mechanics of unsaturated soils. The strong atomic interaction across the clay-water interface could cause a divergence from the bulk phase behavior of water. Examples include capillary condensation and solid-water adsorption (e.g., Shi and Dhir, 2009; Leroy and Müller-Plathe, 2010; Scocchi et al., 2011; Botan et al., 2011). To the best of our knowledge, few studies have used MD simulations to investigate soil-water retention curves accounting for water-air interface and soil-water adsorption. In this article, MD is utilized to study the impacts of the water-air interface and soil-water adsorption on the nanoscale soil-water retention mechanism at low degrees of saturation.

The area of the water-air interface (i.e., concave water meniscus) in unsaturated soils is nontrivial to compute from the MD simulation data. In this article, the point cloud method coupled with surface reconstruction, as detailed in Section 2, will be used to calculate the water-air interface area at different degrees of saturation (water mass content). Note that surface reconstruction is a subject in computer graphics that deals with surface/shape properties of a point set, such as surface normal estimation (Boissonnat, 1984; Edelsbrunner and Mücke, 1994). Among various surface reconstruction techniques, the alpha-shape method has been successfully employed to characterize the shape of molecules like proteins (Peters et al., 1996; Liang et al., 1998). In Wilson et al. (2009), the author validated the robustness and effectiveness of the alpha-shape method in characterizing the shapes of small molecules compared to other shape predictors. In Singh et al. (1996), the authors applied the alpha-shape method in molecular recognition and identified binding sites in proteins. Inspired by the broad applications in molecular biology, in this study, the alpha-shape method was utilized to calculate the water-air interface area from the MD results.

Section 2 presents the unsaturated clay model for the MD simulations and the alpha shape method for the interfacial area calculation. Section 3 concerns the numerical results of the water-air interface area, capillary and adsorptive pressures and conducts data analytics regarding SWRC through a machine-learning curve fitting technique. Section 4 compares the nanoscale water retention mechanisms of kaolinite and pyrophyllite and discusses the effect of clay particle configurations and pore sizes on the clay-water retention mechanism, followed by a summary in Section 5.

**2. Material model and MD modeling**

In this study, pyrophyllite that is a 2:1 clay mineral composed of silicon tetrahedral and aluminum octahedral layers was chosen due to its stable structure and being a precursor to other smectite clay minerals. The aluminum octahedral (O) sheet is bounded by two opposing silicon tetrahedral (T) sheets, which form a T-O-T structure.

The chemical formula of pyrophyllite is $Al_2[Si_4O_{10}](OH)_2$. The unit cell of pyrophyllite has the dimensions of 5.28 Å × 9.14 Å × 6.56 Å in the x-y-z Cartesian coordinate system (Skipper et al., 1995). Figure 1 shows one unit cell of pyrophyllite made up of six types of atoms. In Figure 1, ao is aluminum in the octahedral layer, ocl and hcl are oxygen and hydrogen in the octahedral layer that form the covalent hydroxyl bond, sz is silicon in the tetrahedral layer, oss is oxygen in Si-O-Si linkages, and oas is oxygen in Si-O-Al linkages in the tetrahedral layer. Unlike other smectite clay minerals with a strong cation exchange capacity, pyrophyllite has a weak capacity to swell or shrink because of its neutral surface charge. Considering its structural stability, the clay particle is treated as a rigid body, and its motion was frozen during the simulation. Figure 2 shows the initial configuration of the clay-water model. Each clay layer consists of 36 × 21 × 1 unit cells in the *x-y-z* directions, corresponding to 190.08 Å × 191.94 Å in the x-y plane. Water is modeled by the TIP3P model (Jorgensen et al., 1983). Water molecule is considered rigid body during the simulation. The space between the parallel clay layers is $d = 50$ Å, which could avoid any possible interlayer interactions between clay plates (Amarasinghe et al., 2014).

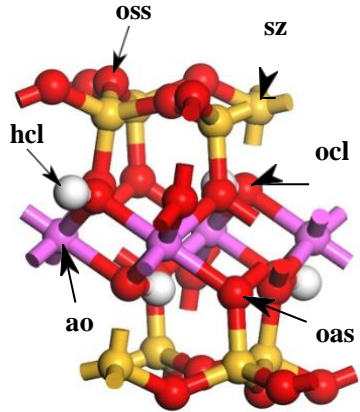

Figure 1: Unit cell of pyrophyllite.

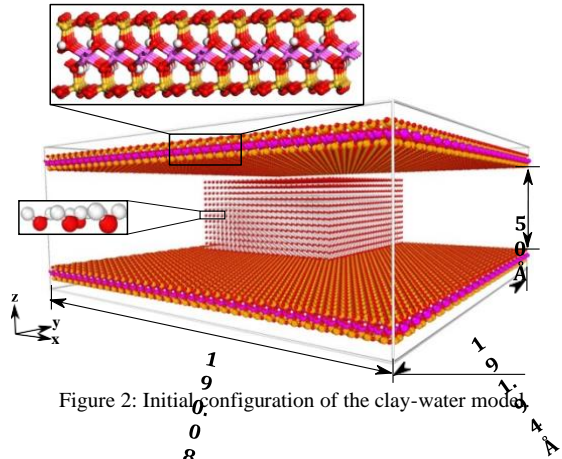

Figure 2: Initial configuration of the clay-water model.

In this study, CHARMM force field (Brooks et al., 1983) is employed to describe soil-water interaction in that CHARMM force field is compatible with the TIP3P water model (Berendsen et al., 1987). CHARMM force field has been widely utilized to study clay-water systems (Katti et al., 2015; Song and Wang, 2019). The non-bonded potential energy $U$ in the CHARMM force field can be defined as

$$U = \sum_{i \neq j} 4\varepsilon_{ij} \left[ \left( \frac{R_{ij}}{r_{ij}} \right)^{12} - 2 \left( \frac{R_{ij}}{r_{ij}} \right)^{6} \right] + \sum_{i \neq j} \frac{q_i q_j}{4\pi\varepsilon_0 r_{ij}}, \quad (1)$$

where $\varepsilon_{ij}$ is the well-depth of Lennard-Jones (LJ) potential and $R_{ij}$ is the distance at the minimum LJ interaction energy, $q_i$ and $q_j$ are the charge of atoms $i$ and $j$, respectively, $\varepsilon_0$ is the vacuum permittivity, $r_{ij}$ is the distance between atoms $i$ and $j$. In this study, clay particles are assumed immobilized and only non-bonded interactions between clay and water were simulated (Song et al., 2018).

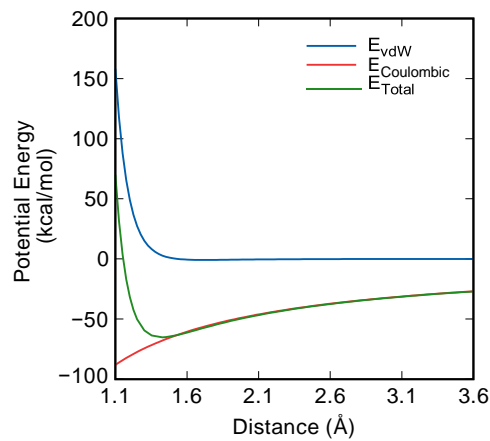

Figure 3: Interactions between clay surface oxygen (oss) and water hydrogen (hw) as a function of the interatomic distance.

The cutoff radii for van der Waals and Coulombic interactions are 10 Å. Table 1 lists the nonbonded parameters for the clay-water system. Note that hw and ow denote water hydrogen and water oxygen, respectively. Figure 3 presents an example of nonbonded interactions (i.e., van der Waals energy and Coulombic energy) between clay surface oxygen (oss) and water hydrogen (hw) as a function of interatomic distance. The interaction between oss and hw was chosen as an example because of the strong clay-water interaction such as surface hydration between the two types of atoms. The minimum potential energy occurs when the two atoms are at a distance of 1.4 Å. This indicates that the forces between the two atoms are repulsive within the distance of 1.4 Å.

Table 1: Values of the input parameters for CHARMM force field.

| Symbol | $q_i$ (e) | $\varepsilon_i$ (kcal/mol) | $R_i$ (Å) |
|---|---|---|---|
| hw | 0.417 | 0.046 | 0.44 |
| ow | -0.834 | 0.152 | 3.53 |
| ocl | -0.96 | 6 | 2.8 |
| hcl | 0.4 | 0.0001 | 2.4 |
| oas | -0.91 | 6 | 2.8 |
| oss | -0.7 | 1 | 3 |
| sz | 1.4 | 0.001 | 7.4 |
| ao | 1.68 | 0.15 | 6.3 |

All MD simulations were performed on LAMMPS, a large-scale atomic/molecular massively parallel simulator (Plimpton, 1995) using NVT ensemble at 298 K. Periodic boundary conditions were assigned in all directions. Water molecules were kept rigid using the SHAKE algorithm (Ryckaert et al., 1977) with constraints applied to hydrogen bonds and angles. The velocity Verlet algorithm with a time step of 0.5 fs (1 fs = $1 \times 10^{-15}$s) was employed to integrate the equations of motion of water. The simulation was first run for 2 ns to bring the system into equilibrium. Then the production simulation was run for 1 ns to output averaged water trajectories and water pressure. The potential energy profile was monitored to check the equilibrium state. Figure 4 shows the time evolution of potential energy for the clay-water system during the equilibration. It can be seen that the system reached a dynamic equilibrium after 0.6 ns.

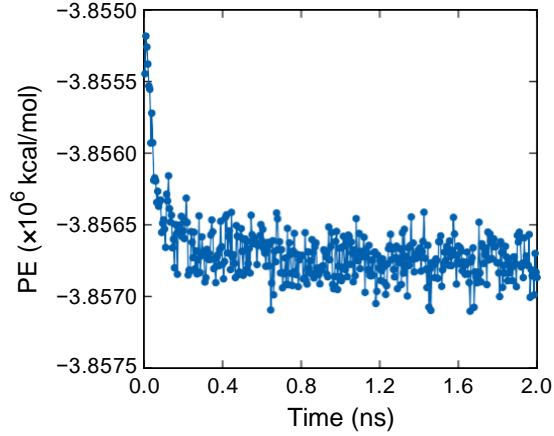

Figure 4: Variation of potential energy during the MD simulation.

In this study, mass water content or moisture content ($\theta_g$) was chosen to represent the degree of water saturation. Mass water content is defined as the ratio between the mass of water and the mass of the solid.

$$\theta_g = \frac{N_w M_w}{N_x N_y N_z M_p}, \tag{2}$$

where $N_w$ is the total number of water molecules, $M_w \approx 18$ g/mol is the molar mass of water, $M_p \approx 360$ g/mol is the molar mass of pyrophyllite, and $N_x = 36$, $N_y = 21$, and $N_z = 1$ are the number of unit cells in the $x$, $y$ and $z$ directions, respectively. The variation of mass water content was realized by adjusting the number of water molecules between the clay particles. The degree of saturation is calculated by

$$S_r = \theta_g G_s / e, \tag{3}$$

where $e$ is void ratio and $G_s$ is specific gravity of dry clay. In molecular dynamics, the pressure tensor of a group of atoms can be expressed through the virial stress tensor (Clausius, 1870) as

$$\sigma_{ij} = \frac{\sum_k^N m_k v_{ki} v_{kj}}{V} + \frac{\sum_k^N r_{ki} f_{kj}}{V}, \tag{4}$$

where $k$ is the atom index, $N$ is the number of atoms in the water group, $V$ is the volume of confined water, $i = j = 1, 2, 3$, and $m_k$, $v_k$, $r_k$ and $f_k$ denote the mass, velocity, position, and force of atom $k$, respectively. The pore water pressure can be determined from equation (4) as

$$p_w = \frac{1}{3}(\sigma_{11} + \sigma_{22} + \sigma_{33}). \tag{5}$$

Next, the method for determining the water-air interface area will be introduced.

*2.1. Procedure of the interfacial area calculation via the alpha-shape method*

In this study, the point cloud concept and alpha shape method were adopted to determine the water-air interface area. Point clouds are commonly produced by 3D scanners, which collect points on the external surfaces bounding the objects. In this work, the MD simulator functions as a 3-D scanner. Figure 5 illustrates the construction of a water point cloud based on trajectories of water molecules. Once the clay-water system reaches equilibrium, we can obtain time-averaged trajectories of water molecules and convert the water body into a 3-D point cloud using the center-of-mass method (Song and Zhang, 2021). The coordinate of center-of-mass $r_c$ of a water molecule can be expressed as

$$r_c = \frac{\sum m_o r_o + m_h(r_{h1} + r_{h2})}{m_o + 2m_h}, \tag{6}$$

where $m$ and $r = (x, y, z)$ denote the atomic mass and Cartesian coordinate system vector, and subscripts $o$ and $h$ stands for water oxygen and hydrogen, respectively.

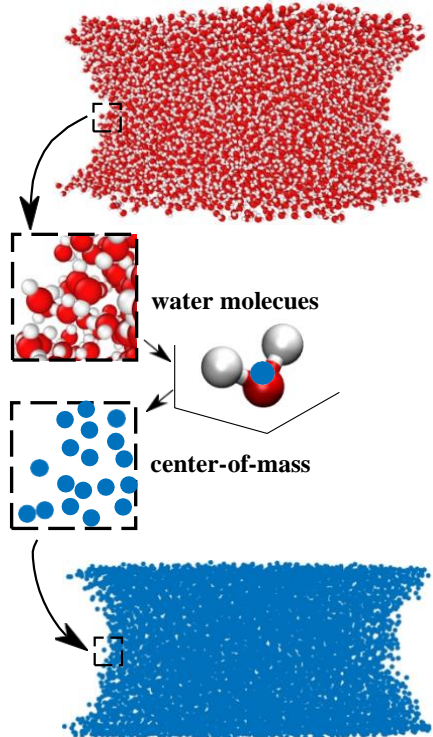

Figure 5: Construction of the water point cloud by extracting the center-of-mass of water molecules.

After obtaining the water point cloud, we reconstruct its surface using the alpha-shape method (Edelsbrunner and Mücke, 1994). The general idea is to find piece-wise triangles (the so-called alpha shapes) to represent the surface of the water point cloud. Figure 6 shows the schematic procedure of interfacial area calculation using the alpha-shape method. We could define a point set $A$ where each point represents the center-of-mass of one water molecule. Let $B$ be an open ball with radius $\alpha$. We restrict $B$ to be empty such that it can occupy its space without enclosing any of the points of $A$, i.e., $B \cap A = \emptyset$. Let $T = \{S_i, S_j, S_k\}$ be a subset of $A$. Given $T$, we could define a 2-simplex $\Delta_T$ (i.e., a triangle) as the convex hull of $T$. Here, the convex hull is the smallest convex set that contains all points in $T$ (Barber et al., 1996). The 2-simplex is said to be $\alpha$-exposed if $T = \partial B \cap A$. Here $\partial B$ is the boundary of $B$.

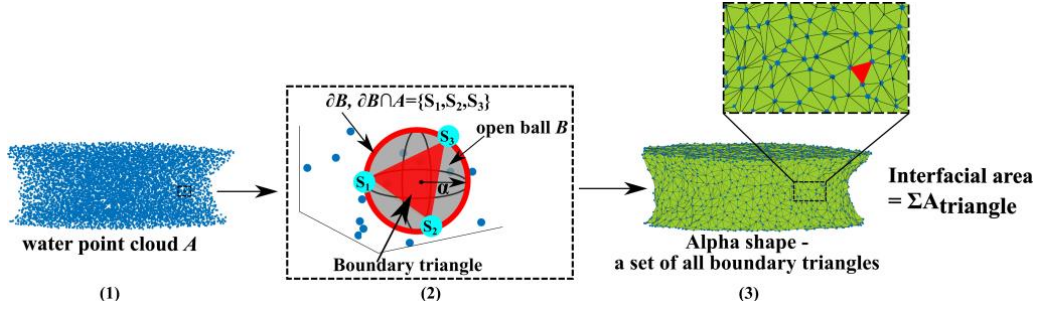

Figure 6: Schematic procedure of the interfacial area calculation using the alpha-shape method.

For implementation, we first perform the Delaunay triangulation of the surface of the water point cloud and then define the alpha complex that associates Delaunay triangulation with the alpha shape (Edelsbrunner and Mücke, 1994). The algorithm adopted is summarized as follows.

a) Compute the Delaunay triangulation of $A$, knowing that the boundary of $\alpha$-shape is contained in Delaunay triangulation.

b) Determine $\alpha$-complex by inspecting all simplices $\Delta_T$ in Delaunay triangulation. If the circumsphere of $\Delta_T$ is empty and the radius of the circumsphere is smaller than $\alpha$, we accept $\Delta_T$ as a member of $\alpha$-complex.

c) All simplices on the boundary of $\alpha$-complex form the $\alpha$-shape.

In Figure 6 (2), the gray object is an empty open ball $B$, the red sphere is the boundary $\partial B$, and $T = \partial B \cap A = \{S_1, S_2, S_3\}$. Meanwhile, we must have $B \cap A = \emptyset$ and $B$ is exterior to $A$. The spherical cap is straightened by a 2-simplex (i.e., triangle in red) connected by points $S_1$, $S_2$, and $S_3$. Thus, the alpha shape of $A$ is the polytope whose boundary consists of all the 2-simplices/triangles. The interfacial area is the sum of the area of each boundary triangle. The parameter $\alpha$ controls the desired level of shape detail. We note that the critical $\alpha$ calculated by MATLAB is assumed as the value of parameter $\alpha$ in this study. It is the smallest alpha that produces an alpha shape with no inner cavities developed.

Figure 7 compares the configurations of alpha shapes when $\alpha$ equals 2 Å, 5 Å, and 8 Å. Alpha smaller than the critical value produces cavities and unreasonable interfacial area. Figure 8 shows parameter ($\alpha$) sensitivity of the wetting-nonwetting interfacial area. The results correspond to a mass water content of 29.1%. The critical alpha for this case is 4.9 Å. It can be found that $\alpha$ has little to no effect on the interfacial area when $\alpha$ is greater than 5 Å. Thus, we assume $\alpha = 5$ Å in this study.

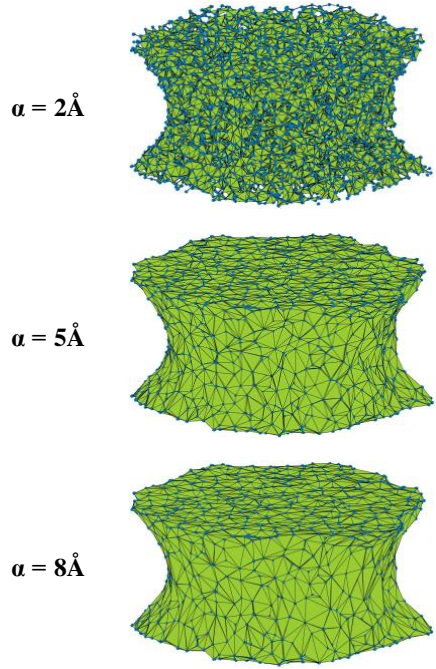

Figure 7: Comparison of alpha shapes of the water point cloud with α =2 Å, 5 Å, and 8 Å.

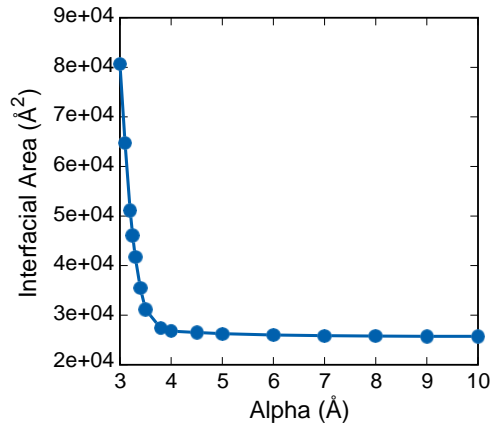

Figure 8: Effect of the alpha value on total interfacial area.

## 3. Numerical results

In this section, we present the numerical results of capillary and adsorptive pressures, the water-air interface area and thickness at different mass water contents, and conduct data analytics regarding adsorptive and capillary water pressure curves through a machine-learning based curving fitting. Table 2 summarizes 24 MD simulations with corresponding values of mass water content and degree of saturation. For each simulation, we computed pore-water pressure (capillary water pressure and adsorptive water pressure), water number density, and interface area and thickness. Through machine learning, we have generated the soil-water retention curve in terms of matric suction, mass water content, and apparent water-air interface area without prescribing a specific functional relationship among those variables. To demonstrate the usefulness of machine-learning based data analytics, the trained and validated neural network was used to predict the matric suction given a mass water content that is beyond the range of the MD simulations in this study.

Table 2: Summary of the MD simulations in this study and their corresponding values of mass water content and degree of saturation.

| Simulation NO. | $\theta_g$ (%) | $S_r$ (%) | Simulation NO. | $\theta_g$ (%) | $S_r$ (%) |
|---|---|---|---|---|---|
| 1 | 6.2 | 2.3 | 2 | 7.3 | 2.8 |
| 3 | 8.4 | 3.2 | 4 | 9.7 | 4.0 |
| 5 | 11.0 | 4.7 | 6 | 12.4 | 5.4 |
| 7 | 13.9 | 6.1 | 8 | 14.7 | 6.3 |
| 9 | 15.5 | 6.9 | 10 | 17.2 | 7.6 |
| 11 | 18.1 | 8.0 | 12 | 19.0 | 8.5 |
| 13 | 20.8 | 9.5 | 14 | 22.7 | 10.5 |
| 15 | 24.8 | 10.6 | 16 | 26.9 | 12.0 |
| 17 | 27.9 | 12.6 | 18 | 29.1 | 13.6 |
| 19 | 31.3 | 14.7 | 20 | 32.5 | 15.1 |
| 21 | 33.7 | 16.0 | 22 | 34.9 | 16.2 |
| 23 | 36.2 | 17.3 | 24 | 37.4 | 17.4 |

*3.1. Adsorptive and capillary water pressures at the nanoscale*

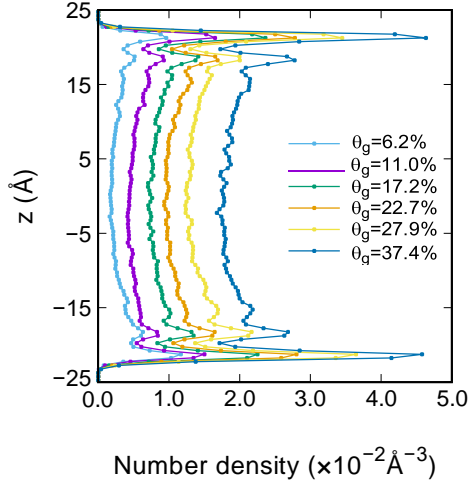

Figure 9: Number density profiles of soil water at various mass water contents.

At the nanoscale, adsorption becomes important in the soil water retention mechanism. In general, the water in the nanoscale clay pore space can be assumed as discrete layers rather than a continuum as the bulk water. Thus, in this study, the water number density profile from the MD results is utilized to distinguish adsorption and capillarity in pore water. First, the nanopore space is evenly divided into a number of parallel layers along the $z$ direction. Then, the number density of each layer is computed as the number of water molecules divided by the volume of the layer. We note that water number density is computed for describing water distribution rather than mass density in that the local mass density of water at the nanoscale deviates substantially from that of bulk water at the macroscale. Figure 9 shows the water number density profiles at various water contents. Despite the different magnitudes of number density, these density curves exhibit a similar mode: the first and second density peaks are located at a distance of 3.75 Å and 6.75 Å from the clay surface (i.e., $z_{clay} = \pm 25$ Å), respectively. Further away from the clay surface, water number density gradually decreases. At the center of the clay nanopore (i.e., $z = 0$), water number density reaches the minimum value. The maximum number density is $2.36 \times 10^{-2}$ Å$^{-3}$ that is about three times the minimum number density (e.g., $0.73 \times 10^{-2}$ Å$^{-3}$ at $\theta_g = 17.2\%$). Large number density fluctuations in the vicinity of the clay-water interface could indicate a strong effect of soil-water adsorption. It also results in a layered water structure. From the number density profile, the soil water could be partitioned into two parts, e.g., the adsorptive water and the capillary water. The adsorptive water layer extends from the clay surface to the trough after the second peak in the number density profile. The capillary region lies in the remaining pore water space where the capillary effect is dominant. The interfaces between the two regions are approximately located at $z = \pm 16.75$ Å from the MD simulation as shown in Figure 9.

Figure 10 plots the variation of percentages of adsorptive and capillary water pressures versus mass water content. Table 3 summarizes the percentages of adsorptive and capillary water pressure in the total pore water pressure at different mass water contents. At a low mass water content, e.g., $\theta_g = 6.2\%$, the adsorptive water pressure occupies 62.9% of the total pore water pressure. As water content increases, the effect of adsorption is gradually weakened. When the mass water content is around 15%, the effects of adsorption and capillarity are similar. The percentage of adsorptive pressure fluctuates around 45% when mass water content exceeds 20%, and capillarity becomes a dominant factor in the overall negative pore water pressure (i.e., matric suction). The MD results have demonstrated that adsorption plays a significant role in the soil water retention mechanism at a low degree of saturation in clay.

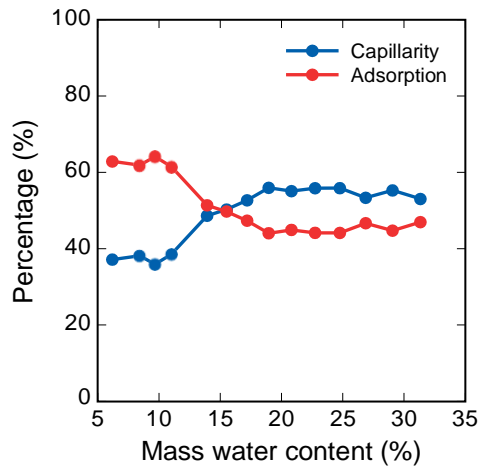

Figure 10: Percentage of adsorptive and capillary water pressure as a function of mass water content.

Table 3: Summary of percentage of adsorptive and capillary water pressures in the pore space for several MD simulations.

| $\theta_g$ (%) | Adsorption (%) | Capillarity (%) |
|---|---|---|
| 6.2 | 62.9 | 37.1 |
| 7.3 | 58.1 | 41.9 |
| 8.4 | 61.9 | 38.1 |
| 9.7 | 64.1 | 35.9 |
| 11.0 | 61.5 | 38.5 |
| 13.9 | 51.4 | 48.6 |
| 15.5 | 49.7 | 50.3 |
| 17.2 | 47.3 | 52.7 |
| 20.8 | 44.9 | 55.1 |
| 24.8 | 44.1 | 55.9 |
| 27.9 | 46.2 | 53.8 |
| 31.3 | 47.0 | 53.0 |
| 34.9 | 42.7 | 57.3 |
| 37.4 | 43.3 | 56.7 |

The adsorptive water pressure from our MD simulations was compared to that from an empirical formula (Tuller et al., 1999) that reads

$$\varphi_{ads} \approx \varphi_{vdW} = \frac{A_H}{6\pi\, t^3}, \qquad (7)$$

where $A_H$ is the Hamaker constant and $t$ is the thickness of adsorptive water layer. For a soil-water system, $A_H$ is on the order of $-10^{-20}$ Joules to $-10^{-19}$ Joules. In this study, it was adopted that $A_H = -6 \times 10^{-20}$ Joules. The results in Figure 9 show that the first non-zero water density occurs near $z = \pm 22.875$ Å, i.e., 2.125 Å away from the clay surface. The thickness of the adsorptive water layer $t$ is approximately 6 Å, e.g., the distance between the trough after the second peak density ($z = \pm 16.875$ Å) and the outermost water layer ($z = 22.875\pm$ Å). From equation (7), we have $\varphi_{ads} = -14.74$ MPa, which is comparable to the adsorptive pressure from our MD results, as illustrated in Figure 22. We note that it lacks experimental testing data to validate the thickness of the adsorptive water layer assumed in our MD simulations.

The pairwise energy and interaction force between the clay and the adsorptive water layer can be used to describe clay-water interactions. Pairwise energy includes the van der Waals component and the long-range Coulombic component. Since only adsorptive water is included, this energy term specifically refers to sorptive energy. Table 4 summarizes the sorptive energy between clay and adsorptive water at different mass water contents. As mass water content increases from 9.7% to 34.9%, the magnitude of sorptive energy increases by 170 % due to increased adsorptive interactions. Figure 11 plots the variation of interaction force with the mass water content. The increase in the adsorptive force is mainly due to the increase of accumulated water molecules in the adsorptive water layer during the wetting process (i.e., increasing water in the pore space).

Table 4: Sorptive energy between clay and adsorptive water at different mass water contents.

| $\theta_g$ (%) | Sorptive energy (kcal/mol) |
|---|---|
| 9.7 | -263.9 |
| 11.0 | -297.1 |
| 12.4 | -328.3 |
| 13.9 | -355.8 |
| 14.7 | -371.3 |
| 17.2 | -388.0 |
| 18.1 | -424.2 |
| 20.8 | -460.8 |
| 27.9 | -597.5 |
| 31.3 | -671.7 |
| 34.9 | -721.5 |

*3.2. Area and thickness of the water-air interface at various mass water contents*

Figure 12 shows a soil-water retention surface in terms of generalized matric suction ($\psi_m$), mass water content ($\theta_g$), and apparent water-air interface area ($A_a$) through a standard curve fitting technique. The water-air interface area was computed through the method introduced in the previous section. Figure 13 shows the water point clouds for 6 simulations. Since clay particles were fixed, the height of soil water remains almost unchanged ($h_w = 43.5 \pm 0.1$ Å) and the soil water body expands along the radial direction with increasing mass water content. We further demonstrate that the mass water contents have a mild effect on the shape of the water meniscus through the contact angle. The contact angle was computed using the method proposed in Song and Zhang (2021). Figure 14 plots the variation of the clay-water contact angle with respect to the mass water content. In Figure 14 "top" and "bot" denote the top and bottom clay-water interfaces, respectively. From the results in Figure 14 it can be postulated that the shape of the water meniscus is almost identical at the different degrees of saturation due to the fixed pore space.

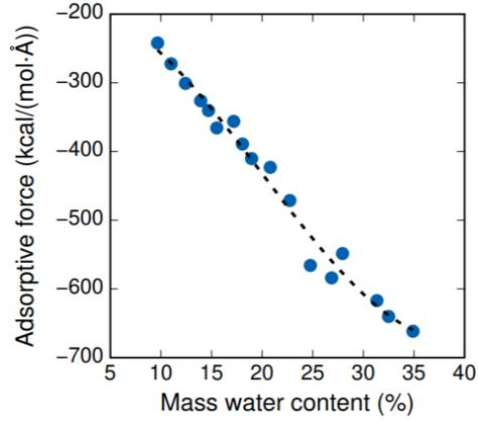

Figure 11: Variation of the adsorptive force with mass water content.

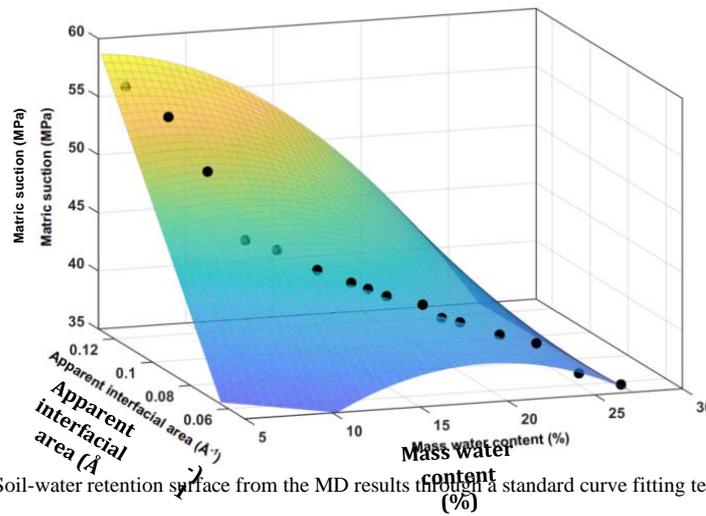

Figure 12: Soil-water retention surface from the MD results through a standard curve fitting technique.

Figures 15 and 16 present the schematic of the water-air interface and the clay-water interface, respectively. Figure 17 plots the water-air interface area versus water mass content. For comparison, the total interface area (i.e., summation of water-air interface area and water-soil interface area) and the water-soil interface area are also plotted. The total interface area increases from 9350 Å$^2$ to 31984 Å$^2$ as mass water content varies from 6.2% to 37.4%. Both the water-air interface area and the clay-water interface area show a nearly linear increase with respect to the mass water content. It can be found that the clay-water interfacial area has a relatively larger growth rate than the water-air interface area with respect to the mass water content. The two curves for water-air and clay-water interface areas intercept at around $\theta_g = 20\%$.

The efficacy of the center-of-mass (COM) method in the surface area calculation was evaluated by comparing the results with the ones from the original (default) water point cloud. The major difference between the two methods is the total number of points in the point cloud. In the default water point cloud, the total number of points is $3N_w$, and each atom determines the coordinate of the corresponding point. In the center-of-mass implementation, the total number of points is $N_w$. Figure 18 compares the water-air interface area calculated from the COM-based point cloud and default point cloud. The deviation is less than 0.1% for all mass water contents.

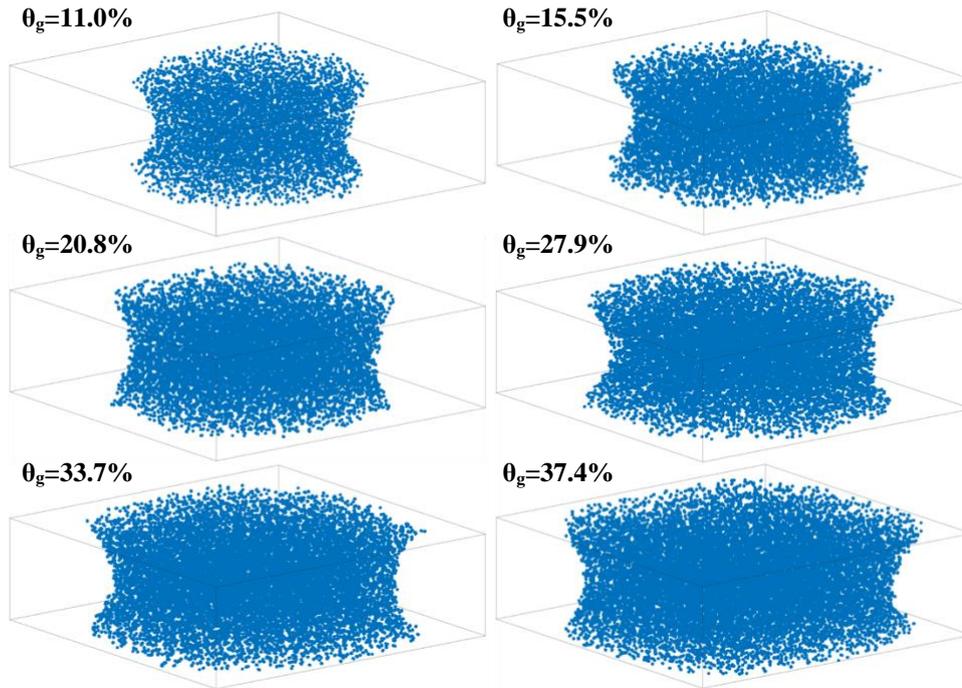

Figure 13: Comparison of water point clouds for six mass water contents.

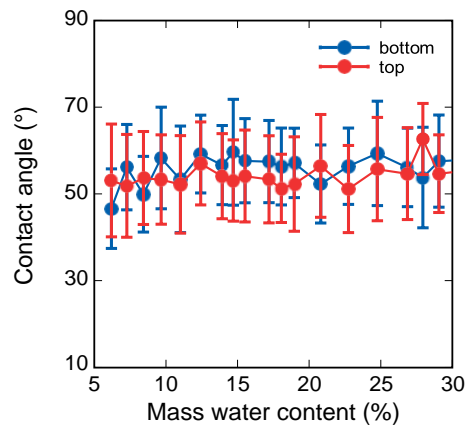

Figure 14: Variation of the clay water contact angle with the mass water content.

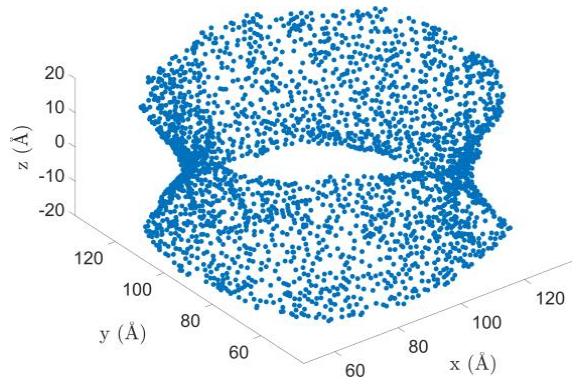

Figure 15: Schematic representation of the water-air interface.

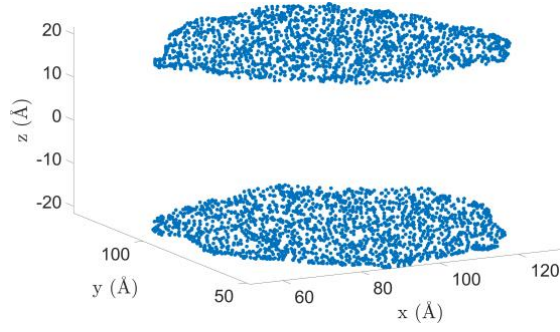

Figure 16: Schematic representation of the clay-water interface.

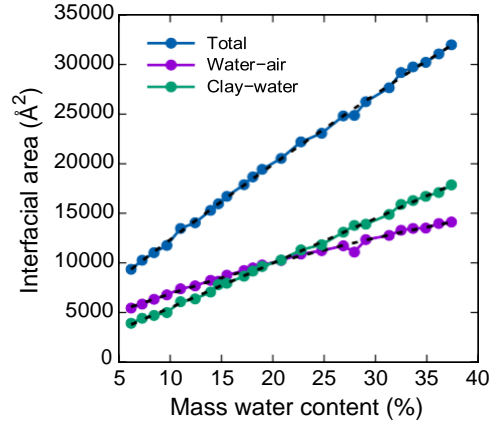

Figure 17: Variations of the water-air interface area and the soil-water interface area versus the mass water content.

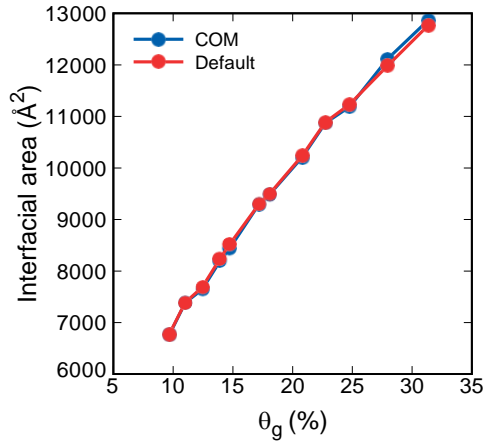

Figure 18: Comparison of the water-air interface area calculated via the center-of-mass (COM) method and the default water point cloud method.

The thickness of the water-air interface is an important physical property of water-air interface (Fredlund, 2006). Based on our MD simulations, we could determine interface thickness from the number density distribution of water in the middle plane of the clay pore. The water between the planes that are parallel to the clay surface at $z = -0.5$ Å and $z = 0.5$ Å were collected and analyzed to avoid the effect of water-clay interactions. Figure 19 presents the number density distribution of water at the pore center along the $x$ and $y$ directions. It can be found that the water number density increases sharply from zero to a peak value near the clay surface. The first density peak is marked with a black circle in Figure 19. For example, a density jump up to 0.036 Å$^{-3}$ can be seen within a distance of $\Delta x = 5$ Å at $\theta_g$=37.4%. Since water density shows a significant change across the water-air interface, we assume that the thickness of the water-air interface is equal to the distance between the outermost water layer and the first density peak. Based on this assumption, the values of interfacial thickness at mass water contents 11.0%, 22.7%, and 37.4% are 5 Å, 5.5 Å, and 5.75 Å, respectively. Previous studies have shown that the thickness of the water-air interface is in the order of 1.5-2 water molecular diameters, e.g., approximately 5 Å (Townsend and Rice, 1991; Fredlund and Rahardjo, 1993; Israelachvili, 2015). This consistency could imply that the water number density from our MD simulations is viable in determining the thickness of the water-air interface in unsaturated clay.

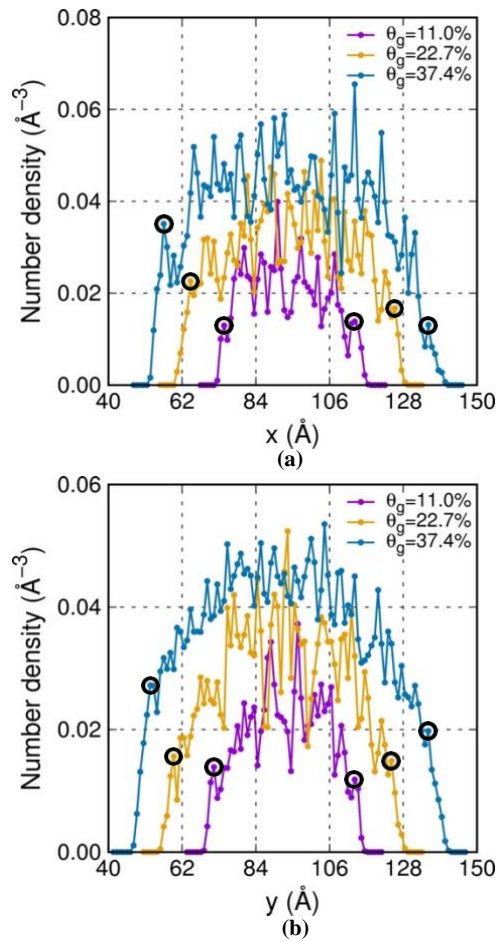

Figure 19: Water number density at the pore center along (a) the *x* - direction and (b) *y* - direction.

*3.3. Soil-water retention curves through neural networks*

In this section, the results of our MD simulation results were analyzed to distinguish adsorption and capillarity in the soil-water retention mechanism. We investigate the relation between matric suction and mass water contents through a neural network (Goodfellow et al., 2016). We first present the results from the MD simulations. Figure 20 plots the variation of matric suction with the apparent interfacial area. The dot represents the mean value, and the error bar represents the standard deviation. Note that the same notations apply to the following figures. In general, matric suction increases with the apparent interfacial area. Here the apparent water-air interfacial area is defined as the water-air interfacial area per water volume. The experimental data on the water-air interfacial area at the nanoscale is not available in the literature. It is noted that the $A_a$-$\theta_g$ relationship obtained from our MD simulations follows the general trend as shown in the results of laboratory tests at the continuum scale (Costanza-Robinson and Brusseau, 2002). Figure 21 plots the variation of matric suction with the mass water content. The results in Figure 21 show that as the mass water content increases from 4.30% to 26.87%, matric suction decreases from 65.4 MPa to 35.86 MPa. Matric suction drops faster as the mass water content is less than 10 %. As the mass water content is greater than 10 %, the decreasing rate of matric suction is lower. Adsorptive water pressure and capillary pressure can be distinguished based on the water density distribution. Figure 22 presents the adsorptive water pressure with the mass water content. The results indicate that the higher the mass water content, the lower the absolute value of adsorptive water pressure. Figure 23 presents the capillary water pressure with the mass water content. The results show that the capillary water pressure oscillates at the average value of -20.71±1.9 MPa. These results are corroborated by the slight variations of contact angles of the clay-water systems at different water contents (See Figure 14).

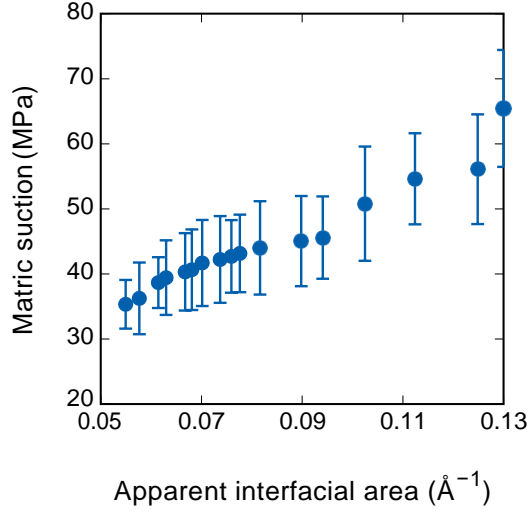

Figure 20: Matric suction versus the apparent water-air interfacial area.

Next, we apply a neural-network deep learning model to predict the relationship between matric suction, the mass water content, and the apparent interfacial area. Given the input (e.g., mass water contents) and the corresponding output (e.g., matric suction), the neural network can generate a model function without any preliminary knowledge about the structure of the function. This is the fundamental difference from the traditional curve fitting of known model functions such as polynomial and power functions. In what follows, we briefly introduce the neural network designed for this study. Figure 24 plots the architecture of the feed-forward neural network (FNN) adopted in this study. With hidden sigmoid neurons and linear output neurons, the neural network allows for fitting 2-dimensional mapping problems. The neural network is trained with the Bayesian regularization al- gorithm. Therefore, the neural network adopted is named by Bayesian regularized feed-forward neural network

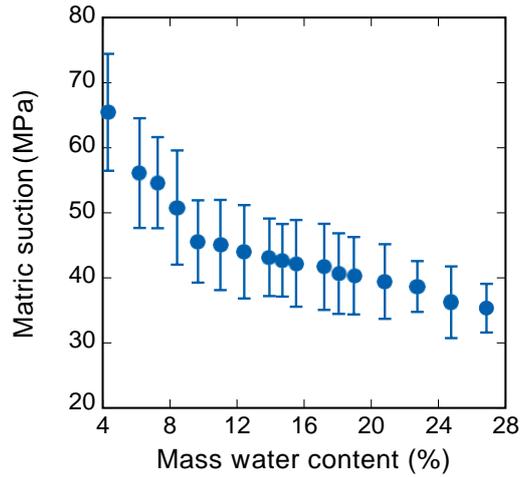
Figure 21: Matric suction versus the mass water content.

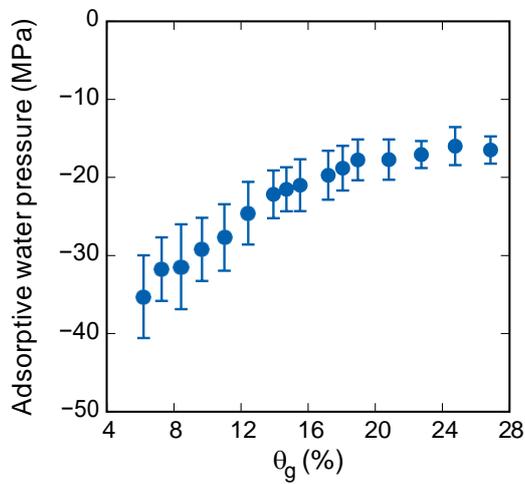
Figure 22: Adsorptive water pressure versus the mass water content and the apparent interfacial area.

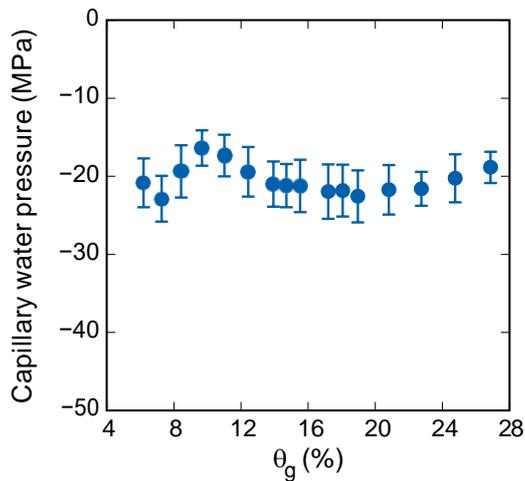
Figure 23: Capillary water pressure versus the mass water content and the apparent interfacial area.

(BRFNN). Bayesian regularization is a mathematical process that converts a nonlinear regression into a "well-posed" statistical problem (MacKay, 1992). The initial weights are assigned by the algorithm in (Nguyen and Widrow, 1990), and the optimization is performed by the Gauss-Newton algorithm (Foresee and Hagan, 1997).

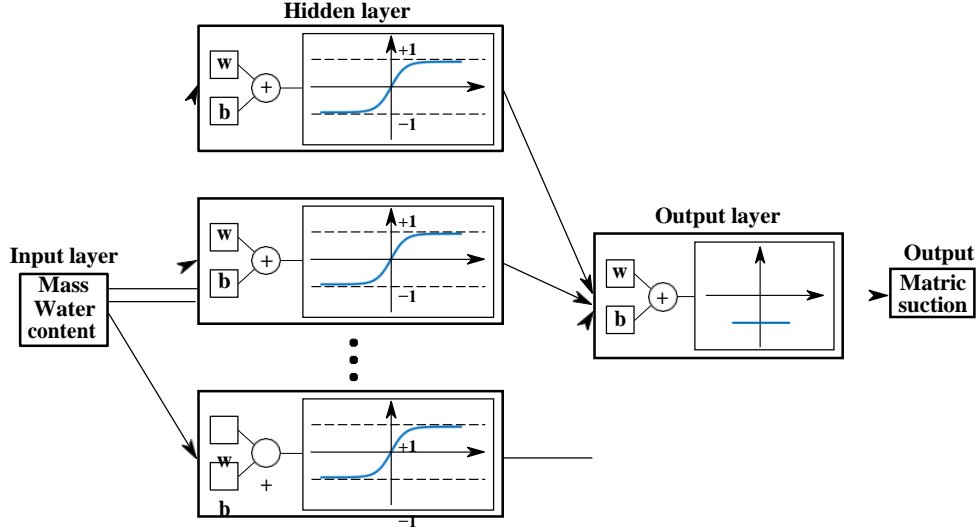

Figure 24: Architecture of the feed-forward neural network.

The Bayesian regularization approach involves the probability distribution of neuron network weights, which differs from conventional network training algorithms (i.e., the optimal weight set chosen by minimizing the error function). Thus, the network predictions are in the form of a probability distribution. The salient feature of BRFNN is that the validation process is not needed (Burden and Winkler, 2009). The input of the neural network is the vector of mass water content. Once received by the hidden layer and multiplication, they are passed to the neurons of the output layer. In the hidden layer, a neuron first computes the weighted sum of input vectors. Then, a constant bias is added to the weighted sum. Finally, the value is fed into the activation function to obtain the output. In the backpropagation, the cost function $E$ to be minimized is

$$E = \mu E_D + \nu E_W, \qquad (8)$$

where $\mu$ and $\nu$ are hyperparameters, $E_D$ is the sum of squared errors, and $E_W$ is the error of the weights. The sum of squared errors reads

$$E_D = \sum_{i=1}^{N_D}(y_i - \hat{y}_i)^2, \quad \hat{y}_i = \sum_{j=1}^{N_N} F_{act}(w_j x_i + b_j) \qquad (9)$$

where $N_D$ is the dimensions of the input vect or, $y_i$ is the target value (e.g., matric suction), $\hat{y}_i$ is the predicted value (e.g., matric suction) as a function of the input value $x_i$ (e.g., mass water content), $N_N$ is the number of neurons, $w_j$ and $b_j$ are the weight and bias corresponding to the $j$-th neuron, respectively, and $F_{act}$ is the activation function in the form of a hyperbolic tangent sigmoid (Vogl et al., 1988). The error of weights $E_W$ is written as

$$E_W = \sum_{j=1}^{N_N} w_j^2. \qquad (10)$$

The input vector of the mass water content and the target vector of matric suction are randomly divided into two groups. For instance, 75% of the input data was used for training, and 25% of the input data was used as an independent test of the neural network generalization.

Figure 25 plots the output and target values of matric suction during training and testing and the errors. Figure 26 plots the mean squared error (MSE) variation versus the number of epochs for the training and test data. Here MSE measures the average squared difference between the predicted values of matric suction and the actual values from MD simulations. The number of epochs is a hyperparameter. An epoch is when all the training data is used at once and is defined as the total number of iterations of all the training data in one cycle for training the machine learning model. As shown in Figure 26, the MSE drops significantly with the initial increase of the epoch number. After the oscillation around epoch 4, the mean squared error continues to decrease. The best training performance is 0.43053 at epoch 22. Figure 27 plots the linear regression coefficient $R$ between the target matric suction and the predicted matric suction for (a) training, (b) test, and (c) the whole dataset. The regression values of the training phase ($R$=0.9951) and the test phase ($R$=0.99976) indicate a good match between the target and the model output.

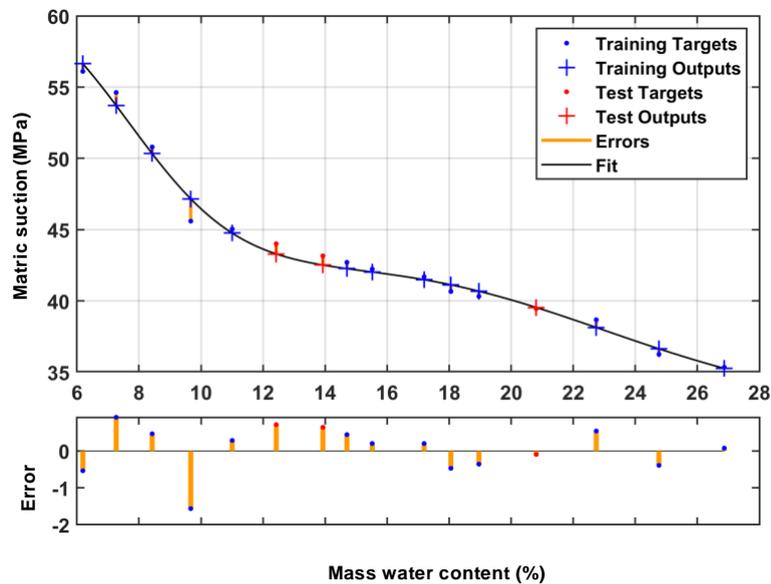

Figure 25: Variation of matric suction versus the mass water content through FNN.

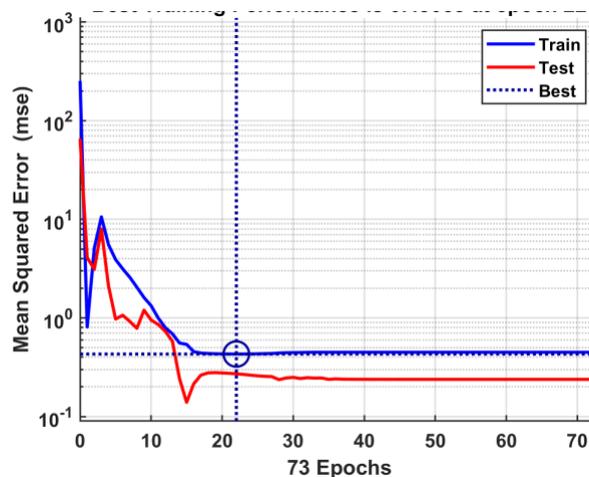

Figure 26: Performance of the training for the relation bewteen matric suction and the mass water content.

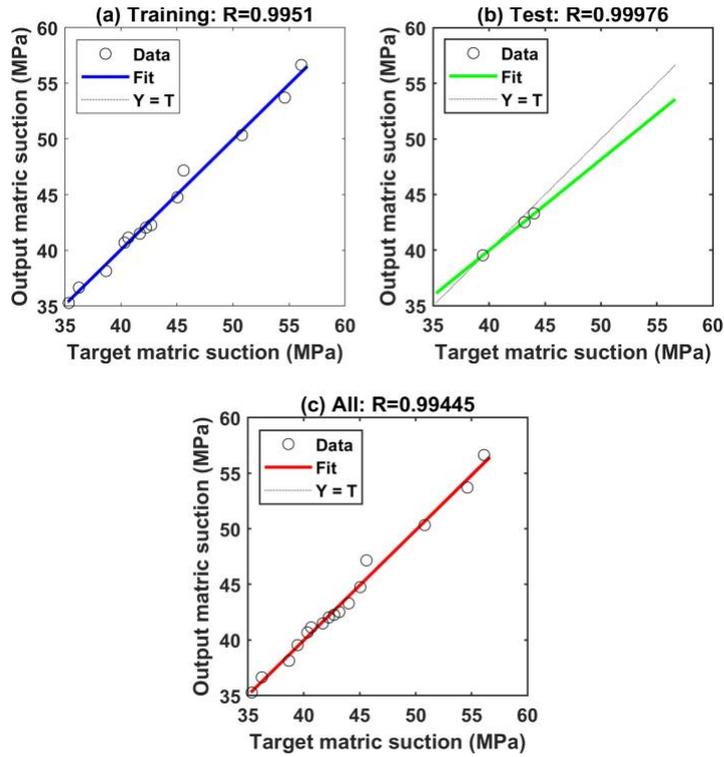

Figure 27: Plot of the regression during training for the curve of matric suction versus the mass water content.

Figures 28 plots the predicted and the target matric suction during training and testing, given the input of the apparent interfacial area. Figure 29 shows the variation of mean squared errors with the epoch number. The results indicate that the best performance is 0.06977, obtained at epoch 51. Figure 30 plots the linear regression coefficient $R$ between the target and the predicted matric suction during (a) training, (b) test, and (c) the total dataset. The regression values of the training phase ($R=0.99915$) and the test phase ($R=0.98939$) indicate a good fit between the target and the model output.

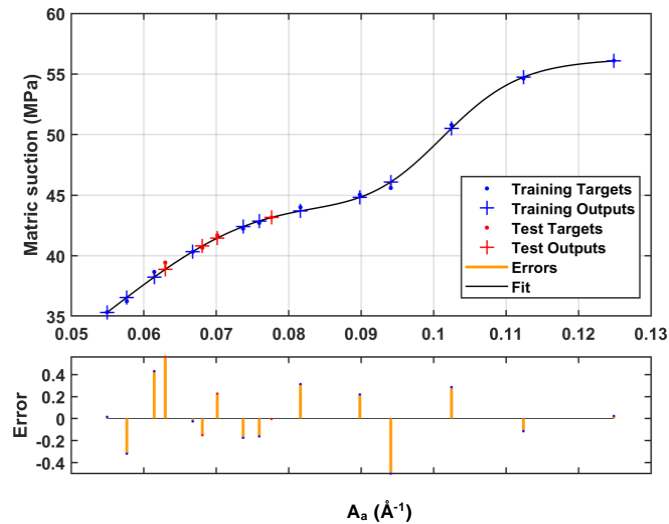

Figure 28: Plot of the variation of matric suction versus the apparent interfacial area through FNN.

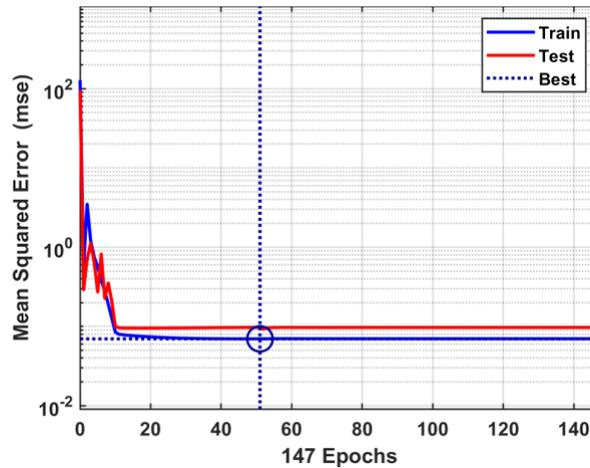
Figure 29: Performance of the training for the curve of matric suction versus the apparent interfacial area.

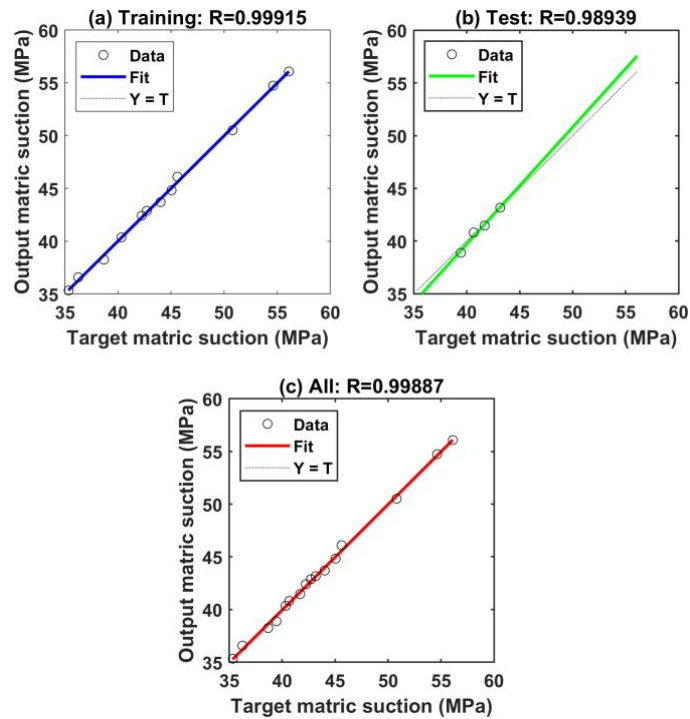
Figure 30: Plot of the regression during the training for the relationship between matric suction and the apparent interfacial area.

Furthermore, we use the trained model to predict matric suction and the apparent interfacial area at mass water contents outside the range of the trained data. Figure 31 shows the variation of matric suction with the mass water content. Figure 32 plots the variation of matric suction and the apparent interfacial area. Combining the results in Figures 31 and 32, Figure 33 presents the relation among matric suction, the mass water content, and the apparent interfacial area. Overall, the results show that neural-networks-based machine learning is a useful tool to analyze the MD results that explicitly consider the soil-water adsorption and generate the soil-water retention curves without prescribing a specific functional relationship between matric suction, the water mass content, and the apparent interfacial area.

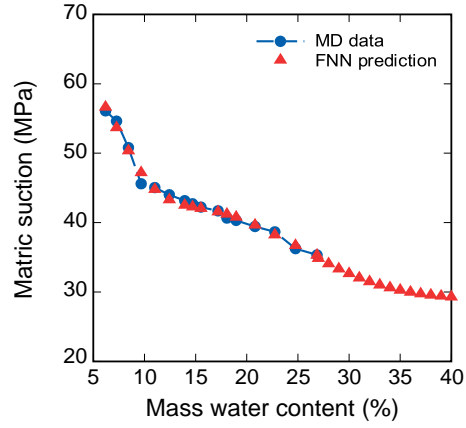
Figure 31: Variation of matric suction with the mass water content from the MD results and the trained FNN.

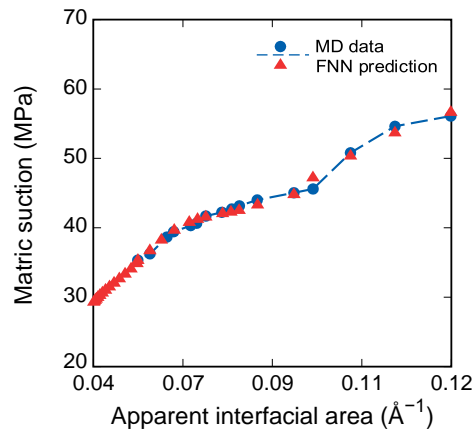
Figure 32: Variation of matric suction with the apparent interfacial area from the MD results and the trained FNN.

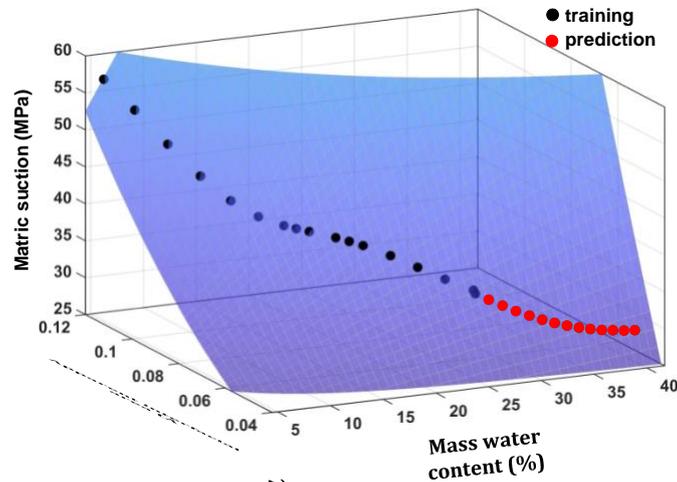
Figure 33: Relationship among matric suction, the mass water content, and the apparent interfacial area.

## 4. Discussions

In this section, we first present the water adsorptive mechanisms in pyrophyllite and kaolinite through density functions, radial distribution functions, and water molecule orientations on the clay surface. Then we discuss the effect of clay particle configurations and pore geometrical sizes on the clay-water adsorption mechanism at the nanoscale through MD simulations with different clay particle configurations and pore geometrical dimensions.

## 4.1. Clay mineral types

It is known that clay mineral types impact the clay-water adsorption mechanism. We compare the water-adsorption mechanism between kaolinite and pyrophyllite at the atomic scale. Kaolinite is a 1:1 type clay mineral. The primary water adsorption mechanism of kaolinite is surface hydroxyl hydration. For comparison, we construct an unsaturated kaolinite-water model with the exact dimensions of the pyrophyllite-water model in Section 2. The mass water content is assumed 22.741%. Figures 34 and 35 compare the number density of water oxygen and water hydrogen in the pyrophyllite pore and the kaolinite pore, respectively. The larger peak density and shorter distance from the first peak to the clay surface demonstrate that kaolinite has a greater water adsorption capacity. This is corroborated by the larger matric suction in the kaolinite-water model than that in the pyrophyllite-water model, i.e., 49.50 MPa versus 38.67 MPa. The radial distribution function (RDF) is used to detect the location of the adsorptive water layer. Figures 36 and 37 plot the RDF of the atom pairs Oc-Ow (i.e., clay tetrahedral oxygen-water oxygen) and Oc-Hw (i.e., clay tetrahedral oxygen-water hydrogen) near the two clay surfaces, respectively. The results show that the larger RDF of water in the kaolinite pore implies greater water absorption at the kaolinite surface. For instance, the distance between the first peak of RDF to the kaolinite surface is about 4 Å which agrees with the location of the first peak water number density as shown in Figures 34 and 35.

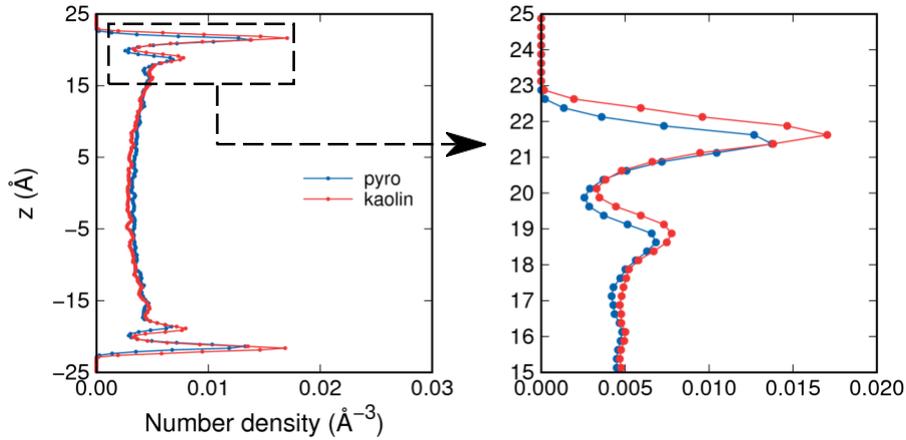

Figure 34: Number density of the water oxygen in the clay pore.

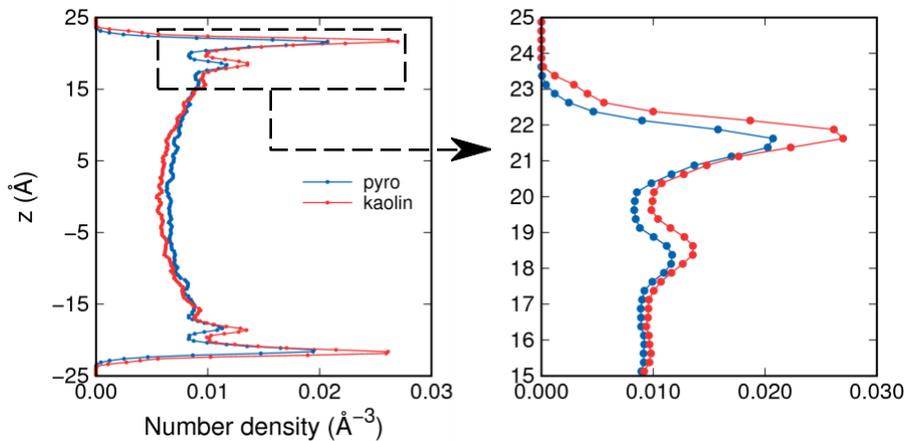

Figure 35: Number density of the water hydrogen in the clay pore.

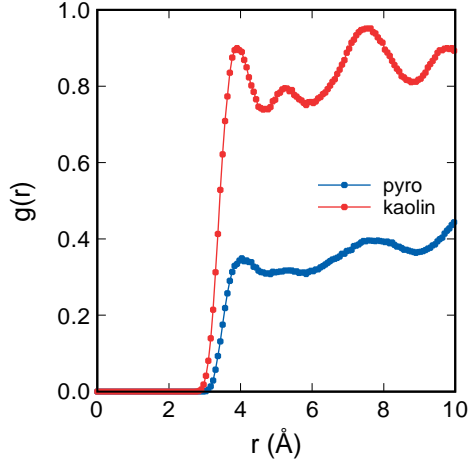

Figure 36: Radial distribution function of the atom pair Oc-Ow near the clay surface.

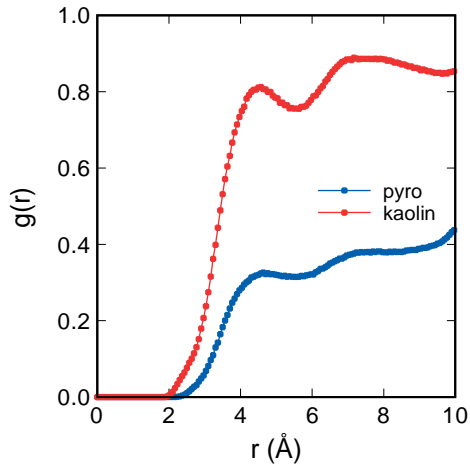

Figure 37: Radial distribution function of the atom pair Oc-Hw near the clay surface.

Furthermore, the snapshots of the MD simulations are presented to show the water molecular distribution due to adsorption at the two clay surfaces. Figure 38 and 39 show the snapshots of the water molecule in the pyrophyllite and kaolinite pores during the adsorptive process at different run times, respectively. The results show that the water molecules move closer to the clay surface and exhibit orientation in both clay pores from 1.5 ps to 10.5 ps due to adsorption. The difference is the distance between the closest water molecule layer to the clay surface. The distances are 1.48 Å for the kaolinite-water and 2.01 Å in the pyrophyllite-water, which indicates the difference in adsorption mechanisms and adsorption strength between kaolinite and pyrophyllite.

*4.2. Clay particle configurations*

The configuration of the two clay platelets could affect the soil water retention mechanism. To study the effect of clay platelet configurations on soil water retention, we simulated three cases where two clay particles are aligned with an angle of 10°, 20°, and 30°, respectively. The three configurations are denoted by cases 1, 2, and 3, respectively. The center-to-center distance between the two clay particles remains the same, i.e., 5 nm. The mass water content is 31.3% for each case. Due to the particle rotation, the dimension of the simulation box in the z-dimension is increased as well. The equilibrated configurations are shown in Figure 40. Using the alpha-shape method, we compute the water-air interface area for the three cases. The total interface areas, i.e., the summation of the soil-water interface area and the air-water interface area, are approximately 28338 $Å^2$, 28242 $Å^2$, and 29035 $Å^2$, respectively. Table 5 summarizes the percentage of the soil-water interface area and the air-water interface area for the three angled clay particle configurations. MD results show the soil-water model with a larger angle between clay particles generates a larger matric suction. The matric suction for cases 1, 2, and 3 are 37.19±5.57 MPa, 39.75±4.74 MPa, 45.83±3.24 MPa, respectively. Table 6 compares the percentage of the adsorptive water pressure and the capillary water pressure under four clay particle configurations. Thus, the adsorptive water pressures at the 3 configurations are -17.26±2.59 MPa, -18.30±2.18 MPa, and -21.75±1.54 MPa, respectively.

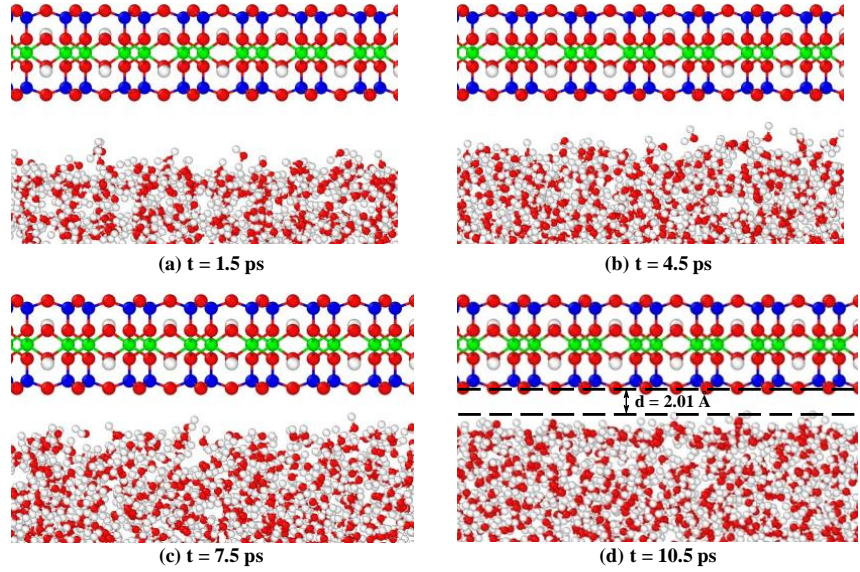

Figure 38: Snapshots of the water molecules in the pyrophyllite pore.

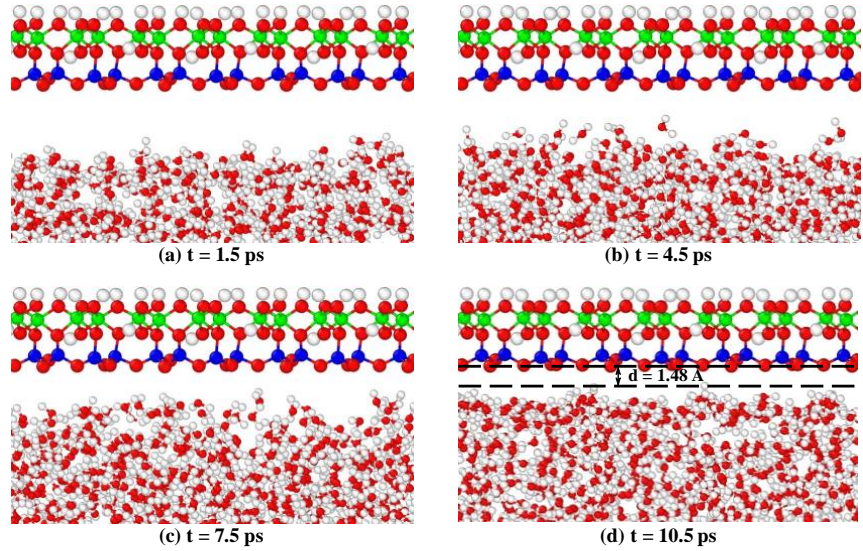

Figure 39: Snapshots of the water molecules in the kaolinite pore.

Table 5: Percentage of the soil-water interface area and the air-water interface area for the three angled clay particle configurations.

| Angle between two clay particles (°) | Clay-water area (%) | Air-water area (%) |
|---|---|---|
| 10 | 48.5 | 51.5 |
| 20 | 50.0 | 50.0 |
| 30 | 55.0 | 45.0 |

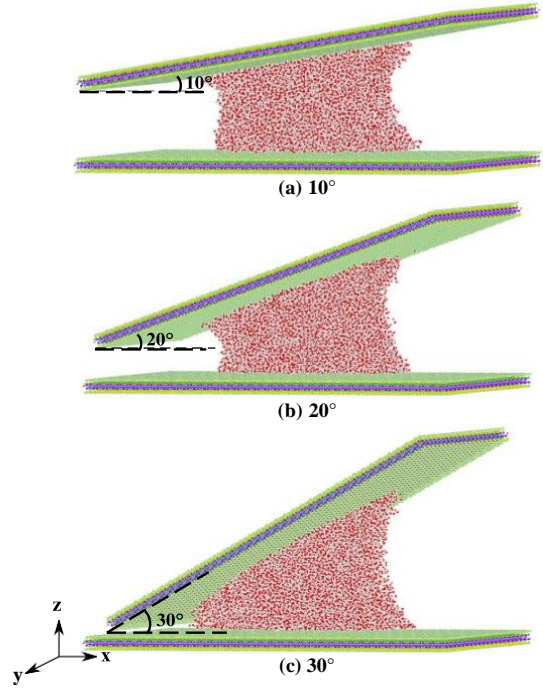

Figure 40: Clay-water models with the two clay particles aligned with an angle at (a) 10°, (b) 20°, and (c) 30°.

Table 6: Percentage of the adsorptive water pressure and the capillary water pressure under four clay particle configurations.

| Angle between the two clay particles (°) | Adsorptive water pressure (%) | Capillary water pressure (%) |
| --- | --- | --- |
| 10 | 46.4 | 53.6 |
| 20 | 46.0 | 54.0 |
| 30 | 47.5 | 52.5 |

*4.3. Effect of the pore width*

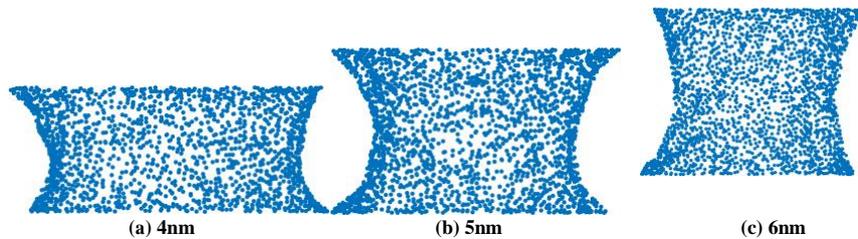

Figure 41: Configurations of equilibrated soil water in the clay pore of width: (a) 4 nm, (b) 5 nm, and (c) 6 nm.

In this part, we investigate the effect of pore width, i.e., the distance between the two clay particles, on matric suction and the interfacial area. We compare the results from the three pores widths, i.e., 4 nm, 5 nm, and 6 nm, under the mass water content of 15.1%. Table 7 summarizes the results. It is found that the matric suction decreases with increasing pore width. This is partially due to the reduced capillary force with increasing pore width. The water-air interface area rises with the pore width. To interpret this, we compare the point clouds of soil water molecules for the three cases, as shown in Figure 41. The radii of the clay-water interface are 7.9 nm, 7.4 nm, and 6.9 nm for the three pores, respectively. However, the change in the pore width might be a dominant factor in that the case for a larger width generates a larger water-air interface area under the same conditions.

Table 7: Summary of matric suctions and water-air interface areas for three pore widths under the same water mass content.

| Pore width (nm) | Matric suction (MPa) | Water-air interface area ($\text{Å}^2$) |
|---|---|---|
| 4 | 46.34±5.69 | 7659 |
| 5 | 42.23±6.68 | 8762 |
| 6 | 39.06±4.09 | 9930 |

## 5. Concluding remarks

We have conducted MD simulations to investigate the soil-water adsorptive and capillary mechanisms of unsaturated clay. The MD model consists of two parallel clay plates and water confined in the clay nanopore. MD simulations were performed at low mass water contents. For processing the MD results, soil water was represented by a point cloud through the center-of-mass method. The water-air interfacial area was calculated using the alpha-shape method. Adsorption is explicitly considered by distinguishing adsorptive pressure from capillary pressure. We have characterized the adsorptive water layer based on the water density profile. The thickness of the adsorptive water layer and the adsorptive water pressure from our MD results are consistent with the results in the literature. For the first time, the feed-forward neural network that does not require a prior function was utilized to generate the nanoscale soil-water retention curve in terms of matric suction, the mass water content, and the apparent interfacial area. The application of the trained neural network was demonstrated by predicting matric suction beyond the range of trained mass water contents. The MD results have demonstrated that adsorption is a dominant mechanism of the nanoscale soil water retention under a low mass water content. For instance, the adsorptive water pressure accounts for more than 60% of the total pore-water pressure at the low mass water content in this study. We note that the study in this article is limited to the pore between two clay particles. The assemblage of clay particles should be considered to study the nanoscale clay-water retention mechanism through MD, which is ongoing research and will be reported in a future publication.

**Acknowledgments**

This work has been supported in part by the US National Science Foundation under contract numbers 1659932 and 1944009. The support is gratefully acknowledged. Any opinions or positions expressed in this article are those of the authors only and do not reflect any opinions or positions of the NSF. We thank the two anonymous reviewers for their expert review of the original version of this article.

**Conflict of interest**

The authors declare that they have no conflict of interest.

**Data availability statement**

The data in this article is available upon reasonable request by contacting the corresponding author.